\documentclass[
  aps,
  reprint,
  superscriptaddress,
  amsmath,
  amssymb
]{revtex4-2}

\usepackage[T1]{fontenc}
\usepackage[utf8]{inputenc}
\usepackage[english]{babel}
\usepackage{graphicx}
\usepackage{hyperref}
\usepackage{braket}
\usepackage{lipsum}
\usepackage{booktabs,tabularx}

\AtBeginDocument{\renewcommand{\selectlanguage}[1]{}}

\begin{document}

\title{Quantum Kernels are Spectral Tensor Networks}

\author{Erik M. {\AA}sgrim}
\affiliation{KTH Royal Institute of Technology, Stockholm, Sweden}

\author{Stefano Markidis}
\affiliation{KTH Royal Institute of Technology, Stockholm, Sweden}

\date{\today}

\begin{abstract}
Quantum kernels admit Fourier representations whose frequencies are determined by the data-encoding gates of the underlying feature map. We show that entangling tensor kernels are matrix product operator factorizations of the corresponding Fourier coefficient tensors, thereby identifying quantum kernels as spectral tensor networks. By grouping gate-level frequency configurations that yield the same feature-wise frequency, we obtain a grouped Fourier form that induces a more compact spectral tensor network representation of the kernel. We further show that kernel target alignment serves as a bridge between the Fourier and tensor network views. On a grid that resolves the accessible Fourier modes, it becomes the Frobenius cosine similarity between Fourier coefficient tensors. Our numerical experiments show that layered quantum kernels admit accurate representations with small bond dimension, revealing a compressibility governed by correlations between Fourier modes. This compressibility provides a diagnostic of classical representability and of whether kernel evaluation is likely to remain classically tractable.
\end{abstract}

\maketitle
Quantum kernels hide a tensor network in their Fourier spectrum. In quantum machine learning, kernels enter as overlaps of quantum feature states~\cite{havlicekSupervisedLearningQuantumenhanced2019,schuld_quantum_2019}, while their expressive structure can be traced to the frequencies generated by data-encoding gates~\cite{schuld_effect_2021,schuld_supervised_2021}. The Hilbert space view connects kernels to states and measurements, whereas the Fourier perspective exposes the spectral structure of the kernel. Entangling tensor kernels (ETKs) were recently introduced as an alternative representation of quantum kernels, given by contractions of data-dependent product states with a matrix product operator (MPO)~\cite{shinQuantumKernelsLens2026}. We show that kernel target alignment (KTA), originally introduced in machine learning as a measure for comparing and learning kernels~\cite{cristianiniKernelTargetAlignment2001,cortesAlgorithmsLearningKernels2012,scholkopfLearningKernelsSupport2001}, provides an operational way to compare spectral structure in quantum kernels.

\begin{figure*}[t]
    \centering
    \includegraphics[width=.95\textwidth]{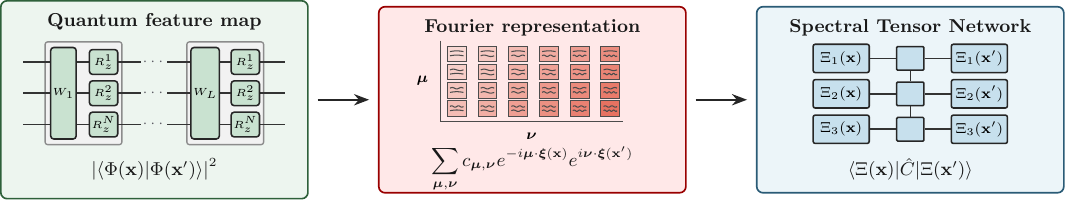}
    \caption{
    From quantum feature maps to spectral tensor networks. A quantum feature map induces a kernel with Fourier coefficients $c_{\boldsymbol{\mu},\boldsymbol{\nu}}$ that encode its spectral structure. Representing this Fourier coefficient object as a matrix product operator yields the spectral tensor network contraction $\langle\Xi(\mathbf{x})|\hat C|\Xi(\mathbf{x}')\rangle$.
    }
    \label{fig:kernel-representations}
\end{figure*}

In this letter, we prove three exact identities. First, the finite Fourier expansion of any quantum kernel can be written directly as an ETK contraction, with the ETK entries given precisely by the Fourier coefficients. Consequently, an exact MPO factorization of the Fourier coefficients gives a spectral tensor network representation of the quantum kernel. Second, repeated feature uploads introduce gate-level frequency degeneracies. Grouping all gate-level configurations that generate the same feature-wise frequency yields an exact representation in terms of feature-wise Fourier modes, and thereby a more compact ETK representation of the kernel. Third, kernel target alignment has an exact coefficient-space form: on an arbitrary data set it is a Fourier-Gram-weighted alignment of coefficient tensors, while on a frequency-resolving grid it reduces to the Frobenius cosine similarity between coefficient tensors. Together, these results identify quantum kernels as spectral tensor networks, as summarized schematically in Fig.~\ref{fig:kernel-representations}. In this representation, the complexity of the kernel is governed by the rank structure of the Fourier coefficient tensor, and low rank signals spectral compressibility.

\newpage
Consider a pure state quantum feature map $\mathbf{x}\mapsto\ket{\Phi(\mathbf{x})}$ and the associated quantum kernel
\begin{equation}
    \kappa(\mathbf{x},\mathbf{x}')
    =
    \left|\braket{\Phi(\mathbf{x})|\Phi(\mathbf{x}')}\right|^2 .
\end{equation}
Although motivated by quantum machine learning, this is more generally a fidelity kernel for a parametrized quantum state~\cite{jozsaFidelityMixedQuantum1994}. The spectral tensor network representation therefore applies beyond supervised learning, whenever repeated evaluations of quantum state overlaps are required. For feature maps in which the data enter through layers of single-qubit data-encoding rotation gates interleaved with data-independent unitaries, the kernel has a finite Fourier expansion~\cite{schuld_effect_2021,schuld_supervised_2021},
\begin{equation}
    \kappa(\mathbf{x},\mathbf{x}')
    =
    \sum_{\boldsymbol{\mu},\boldsymbol{\nu}}
    c_{\boldsymbol{\mu},\boldsymbol{\nu}}\,
    e^{-i\boldsymbol{\mu}\cdot\boldsymbol{\xi}(\mathbf{x})}
    e^{i\boldsymbol{\nu}\cdot\boldsymbol{\xi}(\mathbf{x}')}.
    \label{eq:fourier_kernel_letter}
\end{equation}
Here $\boldsymbol{\xi}(\mathbf{x})=(\xi_1(\mathbf{x}),\dots,\xi_{N_{\rm enc}}(\mathbf{x}))$ collects the classical functions appearing in the individual data-encoding gates. The multi-indices $\boldsymbol{\mu}$ and $\boldsymbol{\nu}$ label the corresponding gate-level Fourier modes. More precisely, each local index $\mu_a$ or $\nu_a$ arises as an eigenvalue difference of the generator of encoding gate $a$. For the single-qubit Pauli-rotation encodings considered here, $\mu_a,\nu_a\in\{-1,0,1\}$. The coefficients $c_{\boldsymbol{\mu},\boldsymbol{\nu}}$ define the spectral object illustrated in Fig.~\ref{fig:kernel-representations}.

\begin{figure}[ht]
    \centering
    \includegraphics[width=0.8\columnwidth]{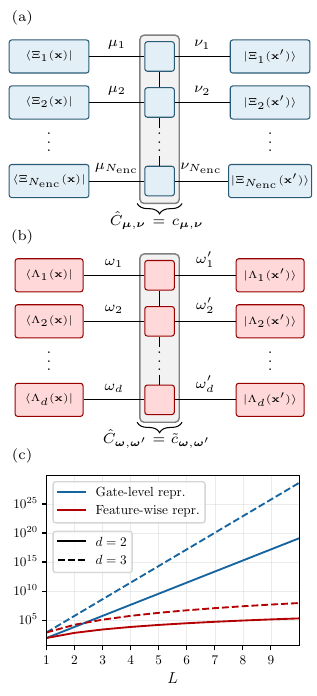}
    \caption{
    Reduction from gate-level to feature-wise frequencies. \textbf{(a)} Gate-level spectral tensor network with $\hat C_{\boldsymbol{\mu},\boldsymbol{\nu}}=c_{\boldsymbol{\mu},\boldsymbol{\nu}}$. \textbf{(b)} Grouping degenerate gate-level configurations yields the feature-wise coefficients $\tilde c_{\boldsymbol{\omega},\boldsymbol{\omega}'}$, with $\hat C_{\boldsymbol{\omega},\boldsymbol{\omega}'}=\tilde c_{\boldsymbol{\omega},\boldsymbol{\omega}'}$. \textbf{(c)} Number of Fourier coefficients in the gate-level and feature-wise representations, scaling as $3^{2dL}$ and $(2L+1)^{2d}$, respectively, for layered feature maps with one upload per feature in each of the $L$ layers.
    }
    \label{fig:site-vs-freq-resolved}
\end{figure}

The first result is the \emph{Spectral tensor network identity}. The tensor-network representation is obtained by collecting the Fourier coefficients into an operator $\hat C$ and contracting it with product-state Fourier features.
For each data-encoding gate $a$, introduce the data embedding $\ket{\Xi(\mathbf{x})}$ as a product state of local Fourier feature vectors
\begin{equation}
\begin{aligned}
    \ket{\Xi_a(\mathbf{x})}
    &=
    \sum_{\mu_a\in\{-1,0,1\}}
    e^{i\mu_a\xi_a(\mathbf{x})}\ket{\mu_a}
    =
    \begin{bmatrix}
    e^{-i\xi_a(\mathbf{x})} \\
    1 \\
    e^{i\xi_a(\mathbf{x})}
    \end{bmatrix},
    \\[0.5em]
    \ket{\Xi(\mathbf{x})}
    &=
    \bigotimes_{a=1}^{N_{\rm enc}}\ket{\Xi_a(\mathbf{x})}.
\end{aligned}
\end{equation}
Here, the local basis is indexed by $\mu_a\in\{-1,0,1\}$. 
Collecting the coefficients $c_{\boldsymbol{\mu},\boldsymbol{\nu}}$ in Eq.~\eqref{eq:fourier_kernel_letter} into
\begin{equation}
    \hat C
    =
    \sum_{\boldsymbol{\mu},\boldsymbol{\nu}}
    c_{\boldsymbol{\mu},\boldsymbol{\nu}}
    \ket{\boldsymbol{\mu}}\bra{\boldsymbol{\nu}},
\end{equation}
we recover the kernel
\begin{equation}
    \kappa(\mathbf{x},\mathbf{x}')
    =
    \bra{\Xi(\mathbf{x})}\hat C\ket{\Xi(\mathbf{x}')}.
    \label{eq:spectral_tn_identity}
\end{equation}
When $\hat C$ is factorized as an MPO, Eq.~\eqref{eq:spectral_tn_identity} is precisely the ETK contraction introduced in Ref.~\cite{shinQuantumKernelsLens2026}, shown in Fig.~\ref{fig:site-vs-freq-resolved}(a). Thus, the Fourier expansion of the quantum kernel is naturally a tensor-network contraction in frequency space. A detailed derivation of this identity is given in Appendix~\ref{app:spectral-tn-identity}.

This identity gives the MPO of the ETK a spectral interpretation. The MPO factorizes the Fourier coefficient operator $\hat C$, and its bond dimension measures the correlations required to represent the kernel's Fourier modes. This notion of bond dimension differs from the usual tensor network description of quantum states, where bond dimension is associated with entanglement across physical partitions~\cite{vidal_efficient_2003,vidal_efficient_2004,orus_practical_2014}. In the spectral tensor network, the complexity instead belongs to the kernel as a function of two inputs, and can exhibit a low-rank spectral structure even when the feature map acts on an exponentially large Hilbert space. This spectral interpretation also makes the notion of compression precise, since the complexity of an MPO is determined by the ranks across its internal cuts. For a bipartition $A|B$ of the frequency sites, reshape the coefficient operator $\hat C$ as
\begin{equation}
    \hat C_{\boldsymbol{\mu},\boldsymbol{\nu}}
    \mapsto
    C_{(\boldsymbol{\mu}_A,\boldsymbol{\nu}_A),(\boldsymbol{\mu}_B,\boldsymbol{\nu}_B)} .
\end{equation}
The singular values of this matricization are the operator Schmidt coefficients across the cut, and the minimum exact MPO bond dimension is their number of nonzero entries, i.e. the operator Schmidt rank. Truncating to a smaller bond dimension retains only the leading Schmidt coefficients and gives a low-bond-dimension approximation of the quantum kernel whose error is controlled by the discarded Schmidt weight.


The second result is the \emph{Feature-wise tensor representation}. The representation in Eq.~\eqref{eq:spectral_tn_identity} assigns one frequency index to each data-encoding gate. To obtain a more compact description, we now restrict to encodings in which each data-encoding gate carries a single raw input feature, so that $\xi_a(\mathbf{x})=x_{i(a)}$. In this setting, repeated uploads of the same feature generate gate-level frequency degeneracies. Let $\mathcal A_i$ be the set of gate indices encoding feature $x_i$, with cardinality $|\mathcal A_i|$ equal to the number of encoding gates carrying that feature. We define the corresponding feature frequency
\begin{equation}
    \omega_i(\boldsymbol{\mu})=\sum_{a\in\mathcal A_i}\mu_a
\end{equation}
and partition the gate-level indices by the feature frequency they generate. The set $\mathcal{S}_{\boldsymbol{\omega}} = \{\boldsymbol{\mu}:\boldsymbol{\omega}(\boldsymbol{\mu})=\boldsymbol{\omega}\}$ then collects all indices with feature frequency $\boldsymbol{\omega}$, giving the grouped coefficients
\begin{equation}
    \tilde c_{\boldsymbol{\omega},\boldsymbol{\omega}'}
    =
    \sum_{\boldsymbol{\mu}\in\mathcal{S}_{\boldsymbol{\omega}}}
    \sum_{\boldsymbol{\nu}\in\mathcal{S}_{\boldsymbol{\omega}'}}
    c_{\boldsymbol{\mu},\boldsymbol{\nu}} .
\end{equation}
With these grouped coefficients, the kernel can be written exactly as~\cite{schuld_effect_2021}
\begin{equation}
    \kappa(\mathbf{x},\mathbf{x}')
    =
    \sum_{\boldsymbol{\omega},\boldsymbol{\omega}'}
    \tilde c_{\boldsymbol{\omega},\boldsymbol{\omega}'}
    e^{-i\boldsymbol{\omega}\cdot\mathbf{x}}
    e^{i\boldsymbol{\omega}'\cdot\mathbf{x}'}.
    \label{eq:frequency-resolved-kernel}
\end{equation}
Our key observation is that this grouped Fourier form can be realized as a distinct ETK architecture whose local indices run over feature-wise frequencies rather than gate-level ones. For feature $x_i$, the corresponding local embedding is
\begin{equation}
\begin{aligned}
    \ket{\Lambda_i(\mathbf{x})}
    &=
    \sum_{\omega_i=-|\mathcal A_i|}^{|\mathcal A_i|}
    e^{i\omega_i x_i}\ket{\omega_i}
    =
    \begin{bmatrix}
    e^{-i|\mathcal A_i|x_i} \\
    \vdots \\
    e^{i|\mathcal A_i|x_i}
    \end{bmatrix}.
\end{aligned}
\end{equation}
Using these local embeddings, the grouped Fourier representation is realized as an ETK with grouped coefficients $\tilde c_{\boldsymbol{\omega},\boldsymbol{\omega}'}$, as shown in Fig.~\ref{fig:site-vs-freq-resolved}(b).
The grouping of gate-level frequencies produces a first compression step before any low-rank tensor-network approximation is introduced. 
For layered feature maps with $d$ input features and $L$ layers, in which each feature is uploaded once per layer, the gate-level representation carries $3^{2dL}$ coefficients, since each encoding gate contributes an independent local frequency index. The feature-wise representation instead keeps only the net Fourier mode of each input feature, reducing the coefficient count to $(2L+1)^{2d}$. The gate-level count therefore grows exponentially in the number of uploads, whereas the feature-wise count grows only polynomially in $L$ for fixed $d$. Fig.~\ref{fig:site-vs-freq-resolved}(c) illustrates this compression for $d=2$ and $d=3$.

The third result is the \emph{Spectral alignment identity}. For two kernel matrices $K$ and $K_Q$ evaluated on the same data set $\mathcal D$, kernel target alignment is
\begin{equation}
    {\rm KTA}(K,K_Q)
    =
    \frac{\langle K,K_Q\rangle_F}
    {\|K\|_F\|K_Q\|_F},
    \label{eq:kta_kernel_matrix}
\end{equation}
where $\langle A,B\rangle_F={\rm Tr}(A^\dagger B)$ is the Frobenius inner product.
KTA is normally a similarity measure between kernel matrices~\cite{cristianiniKernelTargetAlignment2001,cortesAlgorithmsLearningKernels2012,scholkopfLearningKernelsSupport2001}, and has also been used to optimize quantum kernels~\cite{hubregtsenTrainingQuantumEmbedding2022}.
Here it acquires a spectral interpretation.
Let $K$ be the kernel matrix obtained by evaluating the grouped Fourier representation with coefficient tensor $\tilde C$ on a data set $\mathcal D$, and write $K=\Phi^\dagger \tilde C \Phi$,
where $\Phi_{\boldsymbol{\omega},j}
=
e^{i\boldsymbol{\omega}\cdot\mathbf{x}^{(j)}}$ contains the Fourier features. 
Likewise, let $K_Q=\Phi^\dagger \tilde C_Q \Phi$ be the kernel matrix of a target quantum kernel in the same grouped Fourier basis. The encoded Fourier features induce the empirical metric
\begin{equation}
    G=\frac{1}{|\mathcal D|}\Phi\Phi^\dagger.
\end{equation}
In this notation, kernel target alignment takes the form
\begin{equation}
{\rm KTA}(K,K_Q)
=
\frac{
{\rm Tr}\!\left(
\tilde C^\dagger G \tilde C_Q G
\right)
}{
\sqrt{
{\rm Tr}\!\left(
\tilde C^\dagger G \tilde C G
\right)
{\rm Tr}\!\left(
\tilde C_Q^\dagger G \tilde C_Q G
\right)
}
}.
\label{eq:kta-via-G}
\end{equation}
Thus, in general, KTA compares coefficient tensors through the metric $G$.
Assuming a periodic data domain $\mathcal X=[0,2\pi)^d$, this comparison becomes direct when $\mathcal D$ is chosen as a frequency-resolving grid,
\begin{equation}
\begin{aligned}
\mathcal G_{\Omega}
&=
\left\{
\left(
\frac{2\pi a_1}{M_1},
\dots,
\frac{2\pi a_d}{M_d}
\right)
:
a_j=0,\dots,M_j-1
\right\},
\\
M_j
&\ge 2\omega_j^{\max}+1 ,
\end{aligned}
\label{eq:frequency_resolving_grid}
\end{equation}
where $\omega_j^{\max}=|\mathcal A_j|$ in the repeated-upload setting considered here.
On this grid, discrete Fourier orthogonality gives $G=\mathbb I$, and Eq.~\eqref{eq:kta-via-G} reduces to
\begin{equation}
    {\rm KTA}(K,K_Q)
    =
    \frac{\langle \tilde C,\tilde C_Q\rangle_F}
    {\|\tilde C\|_F\|\tilde C_Q\|_F}.
    \label{eq:kta_spectral_alignment}
\end{equation}
KTA therefore becomes the cosine similarity between Fourier coefficient tensors. The derivation of this identity is given in Appendix~\ref{app:detailed-proofs}.
Optimizing a spectral tensor network by KTA aligns the MPO with the Fourier coefficient tensor of the quantum kernel.
This provides the operational bridge between the Fourier description and the tensor network description used in the numerical results below.

\begin{figure*}[t]
    \centering
    \includegraphics[width=.95\textwidth]{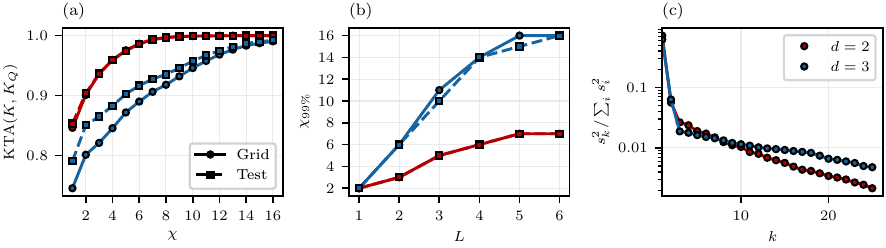}
    \caption{
    Spectral compressibility of quantum kernels. \textbf{(a)} Kernel target alignment between the quantum kernel and its spectral tensor network representation as a function of MPO bond dimension $\chi$ for $L=6$, showing rapid convergence toward unity on both the frequency-resolving grid and independently sampled test points. \textbf{(b)} Minimum bond dimension $\chi_{99\%}$ required to reach average alignment above $0.99$ as a function of depth $L$. \textbf{(c)} Operator Schmidt spectrum of the Fourier coefficient tensor across the first bond for the $L=6$, $\chi=16$ case. The rapid decay of the normalized weights $s_k^2/\sum_j s_j^2$ indicates that only a small number of spectral correlations dominate.
    }
    \label{fig:spectral-compressibility}
\end{figure*}

We use this spectral alignment identity to probe the compressibility of quantum kernels. We consider layered feature maps with one qubit per feature and one upload of each feature per layer, where each layer consists of an entangling block of Haar-random single-qubit unitaries followed by a cyclic CNOT chain and a subsequent layer of data-encoding $R_z$ gates. All reported quantities are averaged over five independently sampled realizations of the random feature map. For each circuit depth, the spectral tensor network is constructed with local data embeddings that realize the same accessible feature-wise frequencies as the quantum kernel. To ensure positive semidefiniteness, the coefficient operator is represented in locally purified form~\cite{wernerPositiveTensorNetwork2016,cuevasPurificationsMultipartiteStates2013} as $\hat C=\hat X^\dagger \hat X$, with $\hat X$ an MPO of bond dimension $\chi$. The MPO entries of $\hat X$ are optimized with Adam~\cite{kingmaAdamMethodStochastic2015} to maximize Eq.~\eqref{eq:kta_spectral_alignment} on the frequency-resolving grid. The resulting kernel is also evaluated on 100 independently sampled test points from the domain $[0,2\pi)^d$. These test points are used to verify that the learned spectral tensor network reproduces the quantum kernel away from the frequency-resolving grid. Further numerical details are given in Appendix~\ref{app:numerical-protocol}.

Fig.~\ref{fig:spectral-compressibility}(a) shows that, for $L=6$ layers, the alignment increases rapidly with bond dimension and approaches unity for both $d=2$ and $d=3$, on both the frequency-resolving grid and independently sampled test points.
Fig.~\ref{fig:spectral-compressibility}(b) shows the minimum bond dimension $\chi_{99\%}$ required to exceed average alignment $0.99$ as a function of depth.
This quantity grows much more slowly than the number of accessible Fourier coefficients and saturates over the largest depths tested.
The origin of this behavior is shown in Fig.~\ref{fig:spectral-compressibility}(c), which plots the operator Schmidt spectrum of the Fourier coefficient tensor $\hat C$ across the first bond. 
The rapid decay of the normalized weights $s_k^2/\sum_j s_j^2$ shows that only a small number of spectral correlations dominate. Together, these results indicate that the effective spectral complexity of these kernels can be much smaller than the number of accessible Fourier coefficients.

The spectral tensor network identity is structural and applies to any quantum kernel admitting the finite Fourier representation in Eq.~\eqref{eq:fourier_kernel_letter}. The numerical evidence for small bond dimension, however, is restricted to the random layered feature maps considered here. Other data encodings, circuit families, or problem-specific feature maps may exhibit slower operator Schmidt decay and therefore require larger bond dimension. Thus, spectral compressibility should be viewed as a diagnostic property of a given quantum kernel, rather than a universal feature of quantum kernels.

This diagnostic has direct consequences for the assessment of quantum advantage in kernel methods. A quantum feature map may generate a large Hilbert space and a large accessible Fourier set, but this does not by itself imply a hard kernel. If the Fourier coefficient tensor has rapidly decaying operator Schmidt spectra, the kernel admits a compact spectral tensor network representation and can be evaluated classically with small bond dimension. Such kernels are therefore unlikely to provide an advantage that relies on the complexity of kernel evaluation alone.
Conversely, slowly decaying spectra identify kernels whose correlations between Fourier modes resist tensor network compression and are stronger candidates for quantum advantage based on kernel evaluation.
Our results therefore suggest that the compressibility of the Fourier coefficient tensor should be treated as a central diagnostic for quantum kernels.
Quantum kernels are spectral tensor networks, and their classical representability is governed by correlations in frequency space rather than only by circuit depth, Hilbert space dimension, or entanglement in the feature state.

\begin{acknowledgments}
We acknowledge support from the Wallenberg Centre for Quantum Technology (WACQT), funded by the Knut and Alice Wallenberg Foundation.
\end{acknowledgments}

\section*{Data availability}
The data and code supporting the numerical findings of this study are available at~\cite{asgrim_spectral_kernel_data_2026} and~\cite{asgrim_spectral_kernel_code_2026}, respectively.

\bibliographystyle{apsrev4-2}
\bibliography{references}

\clearpage
\appendix

\section{Derivation of the spectral tensor network identity}
\label{app:spectral-tn-identity}
We consider an $L$-layer feature map in which the data enter through single-qubit $Z$ rotations interleaved with data-independent unitaries~\cite{schuld_effect_2021,schuld_supervised_2021,shinQuantumKernelsLens2026},
\begin{equation}
    \ket{\Phi(\mathbf{x})} =
    \prod_{l=1}^{L}
    \left(
    \bigotimes_{n=1}^{N}
    R_z^{(n)}(\xi^{(l,n)}(\mathbf{x}))
    W^{(l)}
    \right)
    \ket{0}^{\otimes N}.
\end{equation}
We use the compound index $a=(l,n)$ and denote the total number of data-encoding gates by $N_{\rm enc}=NL$.
The finite Fourier representation of the induced quantum kernel follows from the diagonal action of the data-encoding $R_z$ gates in the computational basis. Expanding the feature-state overlap in computational-basis paths, each data-encoding gate contributes a phase factor of the form
\begin{equation}
    \exp\left[-\frac{i}{2}\xi_a(\mathbf{x})z_a\right],
    \qquad z_a\in\{-1,+1\},
\end{equation}
where $z_a$ is the $Z$-eigenvalue associated with gate $a$ along the chosen path. In the kernel $\kappa(\mathbf{x},\mathbf{x}')=|\braket{\Phi(\mathbf{x})|\Phi(\mathbf{x}')}|^2$, these phases combine with their complex-conjugate counterparts from the bra and ket paths, yielding local frequency indices
\begin{equation}
\left\{
\begin{aligned}
\mu_a &= \frac12 (z_a-\tilde z_a),\\
\nu_a &= \frac12 (z'_a-\tilde z'_a),
\end{aligned}
\right.
\qquad
\mu_a,\nu_a \in \{-1,0,1\}
\end{equation}
where primes indicate the $\mathbf{x}'$ side and tildes the second copy from the modulus square. Equivalently, the multi-indices $\boldsymbol{\mu}=(\mu_a)_{a=1}^{N_{\rm enc}}$ and $\boldsymbol{\nu}=(\nu_a)_{a=1}^{N_{\rm enc}}$ take values in $\{-1,0,1\}^{N_{\rm enc}}$. Following the standard Fourier analysis of quantum models~\cite{schuld_supervised_2021}, one then collects all terms with the same eigenvalue-difference vectors $\boldsymbol{\mu}$ and $\boldsymbol{\nu}$ into coefficients $c_{\boldsymbol{\mu},\boldsymbol{\nu}}$, obtaining the finite Fourier expansion
\begin{equation}
\kappa(\mathbf{x},\mathbf{x}') =
\sum_{\boldsymbol{\mu},\boldsymbol{\nu}\in\{-1,0,1\}^{N_{\rm enc}}}
c_{\boldsymbol{\mu},\boldsymbol{\nu}}\,
e^{-i\boldsymbol{\mu}\cdot\boldsymbol{\xi}(\mathbf{x})}
e^{i\boldsymbol{\nu}\cdot\boldsymbol{\xi}(\mathbf{x}')}.
\end{equation}

The spectral tensor network identity follows directly from this expansion. For each data-encoding gate $a$, define the local Fourier feature vector $\ket{\Xi_a(\mathbf{x})}$
\begin{equation}
    \ket{\Xi_a(\mathbf{x})}
    =
    \sum_{\mu_a\in\{-1,0,1\}}
    e^{i\mu_a\xi_a(\mathbf{x})}\ket{\mu_a}
    =
    \begin{bmatrix}
    e^{-i\xi_a(\mathbf{x})} \\
    1 \\
    e^{i\xi_a(\mathbf{x})}
    \end{bmatrix},
\label{eq:app-local-fourier-feature}
\end{equation}
and let the full data embedding be their product state $\ket{\Xi(\mathbf{x})}=\bigotimes_{a=1}^{N_{\rm enc}}\ket{\Xi_a(\mathbf{x})}$, where the local basis is ordered by $\mu_a=-1,0,1$. Collecting the Fourier coefficients into the operator
\begin{equation}
    \hat C
    =
    \sum_{\boldsymbol{\mu},\boldsymbol{\nu}}
    c_{\boldsymbol{\mu},\boldsymbol{\nu}}
    \ket{\boldsymbol{\mu}}\bra{\boldsymbol{\nu}},
    \label{eq:app-center-operator}
\end{equation}
and using
\begin{equation}
\braket{\Xi(\mathbf{x})|\boldsymbol{\mu}}
=
e^{-i\boldsymbol{\mu}\cdot\boldsymbol{\xi}(\mathbf{x})},
\qquad
\braket{\boldsymbol{\nu}|\Xi(\mathbf{x}')}
=
e^{i\boldsymbol{\nu}\cdot\boldsymbol{\xi}(\mathbf{x}')},
\end{equation}
we obtain
\begin{equation}
    \bra{\Xi(\mathbf{x})}\hat C\ket{\Xi(\mathbf{x}')}
    =
    \sum_{\boldsymbol{\mu},\boldsymbol{\nu}}
    c_{\boldsymbol{\mu},\boldsymbol{\nu}}
    e^{-i\boldsymbol{\mu}\cdot\boldsymbol{\xi}(\mathbf{x})}
    e^{i\boldsymbol{\nu}\cdot\boldsymbol{\xi}(\mathbf{x}')}.
\end{equation}
This reproduces the Fourier expansion of the kernel, and therefore $\kappa(\mathbf{x},\mathbf{x}') = \bra{\Xi(\mathbf{x})}\hat C\ket{\Xi(\mathbf{x}')}$.
When $\hat C$ is represented as a matrix product operator, this contraction is precisely the ETK form introduced in Ref.~\cite{shinQuantumKernelsLens2026}. From this perspective, the MPO of the ETK is a spectral tensor network representation of the Fourier coefficient operator of the quantum kernel.

\section{Feature-wise representation and kernel alignment proof}
\label{app:detailed-proofs}
Next, we derive the feature-wise frequency representation used in the main text. We assume a simplified data encoding in which each data-encoding gate uploads a raw feature, so that $\xi_a(\mathbf{x})=x_{i(a)}$. Let $\mathcal A_i$ denote the set of encoding-gate indices associated with feature $x_i$. For a gate-level frequency label $\boldsymbol{\mu}\in\{-1,0,1\}^{N_{\rm enc}}$ (and likewise for $\boldsymbol{\nu}$), the corresponding feature-wise frequency of feature $x_i$ is $\omega_i(\boldsymbol{\mu})=\sum_{a\in\mathcal A_i}\mu_a$. The accessible feature-wise frequencies become $\Omega_i=\{-|\mathcal A_i|,\ldots,|\mathcal A_i|\}$ and the full accessible spectrum factorizes as $\Omega=\Omega_1\times\cdots\times\Omega_d$.

Several distinct gate-level labels can generate the same feature-wise frequency vector. For each $\boldsymbol{\omega}\in\Omega$, we define the corresponding degeneracy sector $\mathcal S_{\boldsymbol{\omega}} = \left\{\boldsymbol{\mu} : \boldsymbol{\omega}(\boldsymbol{\mu})=\boldsymbol{\omega}\right\}$.
Summing the gate-level coefficients over pairs of such sectors gives the grouped Fourier coefficients
\begin{equation}
    \tilde c_{\boldsymbol{\omega},\boldsymbol{\omega}'}
    =
    \sum_{\boldsymbol{\mu}\in\mathcal S_{\boldsymbol{\omega}}}
    \sum_{\boldsymbol{\nu}\in\mathcal S_{\boldsymbol{\omega}'}}
    c_{\boldsymbol{\mu},\boldsymbol{\nu}} .
    \label{eq:app-grouped-coefficients}
\end{equation}
The kernel then becomes
\begin{equation}
    \kappa(\mathbf{x},\mathbf{x}')
    =
    \sum_{\boldsymbol{\omega},\boldsymbol{\omega}'\in\Omega}
    \tilde c_{\boldsymbol{\omega},\boldsymbol{\omega}'}
    e^{-i\boldsymbol{\omega}\cdot\mathbf{x}}
    e^{i\boldsymbol{\omega}'\cdot\mathbf{x}'} .
    \label{eq:app-frequency-resolved}
\end{equation}
This is the feature-wise spectral representation of the quantum kernel~\cite{schuld_effect_2021,schuld_supervised_2021}. In the layered setting considered here, each feature is uploaded once per layer, so $|\mathcal A_i|=L$ for all $i$. For $d$ features uploaded $L$ times, the gate-level representation contains $3^{2dL}$ coefficients, while the feature-wise frequency representation contains $(2L+1)^{2d}$ coefficients. Collecting the grouped coefficients into an MPO $\hat C$ and defining the local feature-wise embeddings
\begin{equation}
\ket{\Lambda_i(\mathbf{x})}
=
\sum_{\omega_i\in\Omega_i}
e^{i\omega_i x_i}\ket{\omega_i},
\qquad
\ket{\Lambda(\mathbf{x})}
=
\bigotimes_{i=1}^d \ket{\Lambda_i(\mathbf{x})},
\end{equation}
the grouped Fourier representation becomes
\begin{equation}
\kappa(\mathbf{x},\mathbf{x}')
=
\bra{\Lambda(\mathbf{x})}\hat C\ket{\Lambda(\mathbf{x}')},
\end{equation}
which is the spectral tensor network used in the main text.

We now prove the alignment identity. 
Kernel target alignment is a standard measure of similarity between kernel matrices~\cite{cristianiniKernelTargetAlignment2001,cortesAlgorithmsLearningKernels2012,scholkopfLearningKernelsSupport2001}, and has also been used to optimize quantum kernels~\cite{hubregtsenTrainingQuantumEmbedding2022}. 
Let $\mathcal D=\{\mathbf{x}^{(j)}\}_{j=1}^{|\mathcal D|}$ be the set of data points used to evaluate the kernel matrices, and define the Fourier feature matrix
\begin{equation}
    \Phi_{\boldsymbol{\omega},j}
    =
    e^{i\boldsymbol{\omega}\cdot\mathbf{x}^{(j)}} .
\end{equation}
Let $K$ be the kernel matrix associated with the grouped coefficient tensor $\tilde C$ on the data set $\mathcal D$, and let $K_Q$ be the corresponding target quantum-kernel matrix associated with the grouped coefficient tensor $\tilde C_Q$. These matrices can be written as
\begin{equation}
    K=\Phi^\dagger \tilde C\Phi ,
    \qquad
    K_Q=\Phi^\dagger \tilde C_Q\Phi .
    \label{eq:app-kernel-fourier-matrix}
\end{equation}

We define the data-induced Fourier overlap matrix $G=\frac{1}{|\mathcal D|}\Phi\Phi^\dagger$ with the entries given by
\begin{equation}
    G_{\boldsymbol{\omega},\boldsymbol{\omega}'}
    =
    \frac{1}{|\mathcal D|}
    \sum_{\mathbf{x}^{(j)}\in\mathcal D}
    e^{i(\boldsymbol{\omega}-\boldsymbol{\omega}')
    \cdot\mathbf{x}^{(j)}} .
    \label{eq:app-fourier-metric}
\end{equation}
Using Eq.~\eqref{eq:app-kernel-fourier-matrix} and cyclicity of the trace, kernel target alignment can be written in coefficient space as
\begin{equation}
{\rm KTA}(K,K_Q)
=
\frac{
{\rm Tr}\!\left(
\tilde C^\dagger G \tilde C_Q G
\right)
}{
\sqrt{
{\rm Tr}\!\left(
\tilde C^\dagger G \tilde C G
\right)
{\rm Tr}\!\left(
\tilde C_Q^\dagger G \tilde C_Q G
\right)
}
}.
\label{eq:app-kta-metric}
\end{equation}
The factors of $|\mathcal D|$ cancel between numerator and denominator.

Assume now that the domain is $[0,2\pi)^d$ and choose the frequency-resolving grid
\begin{equation}
    \mathcal G_\Omega
    =
    \left\{
    \left(
    \frac{2\pi a_1}{M_1},
    \ldots,
    \frac{2\pi a_d}{M_d}
    \right):
    a_j=0,\ldots,M_j-1
    \right\},
    \label{eq:app-grid}
\end{equation}
with $M_j\ge 2\omega_j^{\max}+1$, where $\omega_j^{\max}=|\mathcal A_j|$ under the simplified encoding.
On this grid,
\begin{equation}
    G_{\boldsymbol{\omega},\boldsymbol{\omega}'}
    =
    \prod_{j=1}^{d}
    \frac{1}{M_j}
    \sum_{a_j=0}^{M_j-1}
    \exp\!\left[
    \frac{2\pi i(\omega_j-\omega_j')a_j}{M_j}
    \right].
    \label{eq:app-grid-orthogonality}
\end{equation}
By discrete Fourier orthogonality, each factor is one when
$\omega_j=\omega_j'$ and zero otherwise. 
Hence $G=\mathbb I$ on $\mathcal G_\Omega$, and Eq.~\eqref{eq:app-kta-metric} reduces to
\begin{equation}
    {\rm KTA}(K,K_Q)
    =
    \frac{
    \langle \tilde C,\tilde C_Q\rangle_F
    }
    {
    \|\tilde C\|_F\|\tilde C_Q\|_F
    } .
    \label{eq:app-spectral-alignment}
\end{equation}
Thus, on a frequency-resolving grid, kernel target alignment is the Frobenius cosine similarity between the grouped coefficient tensors.

\section{Numerical protocol}
\label{app:numerical-protocol}

We use the feature-wise frequency representation of Eq.~\eqref{eq:app-frequency-resolved} to probe the spectral compressibility of kernels induced by layered feature maps with one qubit per feature and one upload per layer. Following circuit-centric quantum classifier ansatzes~\cite{schuldCircuitcentricQuantumClassifiers2020a}, each layer consists of Haar-random single-qubit unitaries followed by a cyclic CNOT chain and a layer of data-encoding $R_z$ rotations; all reported quantities are averaged over five random realizations. For the feature maps considered here, the local data-embedding tensors have dimension $2L+1$, matching the accessible feature-wise spectrum of the target kernel.

Since the quantum kernel is real-valued, we use a real trigonometric local embedding and optimize real-valued MPO entries. In the numerical implementation, the local embedding is
\begin{equation}
\begin{aligned}
    \ket{\Lambda_i(\mathbf{x})}
    =
    \bigl[\,1,\sqrt{2}\cos(x_i),\sqrt{2}\sin(x_i),\ldots, {}& \\
    \sqrt{2}\cos(Lx_i),\sqrt{2}\sin(Lx_i)\bigr]^{\mathsf T}.&
\end{aligned}
\label{eq:app-real-embedding}
\end{equation}
This real trigonometric basis contains the same kernel information, and the factors of $\sqrt{2}$ ensure orthonormality on $[0,2\pi)$.

To ensure that the represented kernel is positive semidefinite, we parametrize the coefficient operator $\hat C$ in locally purified form~\cite{wernerPositiveTensorNetwork2016,cuevasPurificationsMultipartiteStates2013}, $\hat C=\hat X^\dagger \hat X$, where $\hat X$ is represented as an MPO with maximum bond dimension $\chi$ and matching local feature-space dimensions, so that the purification does not impose an additional local low-rank restriction. The tensor entries of $\hat X$ are initialized with small Gaussian noise and optimized by maximizing the spectral alignment objective in Eq.~\eqref{eq:app-spectral-alignment}. We use the Adam optimizer~\cite{kingmaAdamMethodStochastic2015} with learning rate $10^{-2}$. For frequency-resolving grids containing at most $256$ points, the objective is evaluated on the full grid at each optimization step. For larger grids, the grid points are partitioned into batches of $256$ points, and the objective is evaluated batch by batch during optimization.

The reported alignment values are evaluated both on the frequency-resolving grid $\mathcal G_\Omega$ and on 100 independently sampled test points from $[0,2\pi)^d$. The grid points are used in the KTA optimization, while the test points are held out and used only to assess how well the learned kernel extends beyond the discrete frequency-resolving grid. The quantity $\chi_{99\%}$ is defined as the smallest bond dimension for which the average kernel target alignment over the random feature maps exceeds $0.99$. To improve convergence at larger bond dimension, we use a progressive initialization strategy. After optimizing $\hat X$ at bond dimension $\chi$, we embed the optimized entries into an MPO with bond dimension $\chi+1$ and initialize the newly introduced entries with small Gaussian noise. This preserves the spectral correlations learned at smaller bond dimension while increasing the expressive capacity of the MPO. The strategy is analogous in spirit to standard tensor network practice, where the bond dimension is increased progressively to improve the representation~\cite{schollwoeck_density-matrix_2011}.


\end{document}